\documentclass[12pt]{article}
\input epsf
\newcommand{\NPB}[3]{\emph{ Nucl.~Phys.} \textbf{B#1} (#2) #3}   
\newcommand{\PLB}[3]{\emph{ Phys.~Lett.} \textbf{B#1} (#2) #3}   
\newcommand{\PRD}[3]{\emph{ Phys.~Rev.} \textbf{D#1} (#2) #3}

\newcommand{\JHEP}[3]{\emph{JHEP} \textbf{#1} (#2) #3}
\def\dalemb#1#2{{\vbox{\hrule height .#2pt
        \hbox{\vrule width.#2pt height#1pt \kern#1pt
                \vrule width.#2pt}
        \hrule height.#2pt}}}

 \def\bd{\begin{document}} \def\ed{\end{document}}
\def\ds{\documentstyle} \let\fr=\frac \let\bl=\bigl \let\br=\bigr
\let\Br=\Bigr \let\Bl=\Bigl 
\let\bm=\bibitem
\let\na=\nabla
\let\pa=\partial \let\ov=\overline
\def\ie{{\it i.e.\ }} 
\newcommand{\pr}{\paragraph{}}
\newcommand{\be}{\begin{equation}}
\newcommand{\ee}{\end{equation}}
\newcommand{\beba}{\begin{equation}\begin{array}{lcl}}
\newcommand{\eaee}{\end{array}\end{equation}}
\newcommand{\bea}{\begin{eqnarray}}
\newcommand{\eea}{\end{eqnarray}}
\newcommand{\ba}{\begin{array}}
\newcommand{\ea}{\end{array}}
\newcommand{\td}{\tilde}
\newcommand{\norsl}{\normalsize\sl}
\newcommand{\ns}{\normalsize}
\newcommand{\refs}[1]{(\ref{#1})}
\def\simlt{\mathrel{\lower2.5pt\vbox{\lineskip=0pt\baselineskip=0pt
           \hbox{$<$}\hbox{$\sim$}}}}
\def\simgt{\mathrel{\lower2.5pt\vbox{\lineskip=0pt\baselineskip=0pt
           \hbox{$>$}\hbox{$\sim$}}}}

\begin{document}
\thispagestyle{empty}
\rightline{\normalsize\sf hep-ph/0004214}
\rightline{\normalsize CPHT-S044.0400}
\rightline{\normalsize Crete-00-12}
\rightline{\normalsize April 2000}
\vskip 1.0truecm
\centerline{\Large\bf 
A D-brane alternative to unification}
\vskip 1.truecm
\centerline{{\large\bf I. Antoniadis}~$^a$, 
{\large\bf E. Kiritsis~$^b$} and {\large\bf T.N. Tomaras}~$^{b}$}
\vskip .5truecm
\centerline{{\it $^a$Centre de Physique Th{\'e}orique~\footnote{Unit{\'e} mixte du
CNRS et de l'EP, UMR 7644.}, Ecole Polytechnique, 91128 Palaiseau, France}}
\vskip .5truecm
\centerline{\it $^b$Department of Physics, University of Crete, and FO.R.T.H.}
\centerline{\it P.O. Box 2208, 710 03 Heraklion, Crete, Greece}

\vskip 0.5truecm
\centerline{\bf\small ABSTRACT}
\vskip .4truecm
We propose a minimal embedding of the Standard Model spectrum in a D-brane
configuration of type I string theory. The $SU(3)$ color and $SU(2)$ weak
interactions arise from two different collections of branes. The model is neither
grand unified nor supersymmetric but it naturally leads to the right prediction of
the weak angle for a string scale of the order of a few TeV. It requires two Higgs
doublets and guarantees proton stability.

\hfill\break
\vfill\eject

String theory is the only known framework for quantizing gravity. If its
fundamental scale is of the order of the Planck mass, stability of the hierarchy
of the weak scale requires low energy supersymmetry. This framework fits nicely
with the apparent unification of the gauge couplings in the minimal
supersymmetric standard model. 
However, breaking supersymmetry at low energies is a hard problem, which in
string perturbation theory implies a large extra dimension, \cite{bac,k,ant}.
Recently, an alternative approach has been put
forward \cite{add,aadd} in which stabilization of the hierarchy is achieved without
supersymmetry, by lowering the string scale down to a few TeV
\cite{ant,Ant,aadd,twoc,st}. A
natural realization of this possibility is offered by weakly coupled type I string
theory, where gauge interactions are described by open strings whose ends are
confined on D-branes, while gravity is mediated by closed strings in the bulk
\cite{aadd}. The observed hierarchy between the Planck and the weak scales is then
accounted for by two or more large dimensions, transverse to our brane-world, with
corresponding size varied from a millimeter to a fermi.

One of the main questions with such a low string scale is to understand the
observed values of the low energy gauge couplings. One possibility is to have
the three gauge group factors of the Standard Model arising from different
collections of coinciding branes. This is unattractive since the three gauge
couplings correspond in this case to different arbitrary parameters of the
model. A second possibility is to maintain unification by imposing all the
Standard Model gauge bosons to arise from the same collection of D-branes. The
large difference in the actual values of gauge couplings could then be explained
either by introducing power-law running from a few TeV to the weak scale
\cite{ddg}, or by an effective logarithmic evolution in the transverse space in
the special case of two large dimensions \cite{abd}. However, no satisfactory model
built along these lines has so far been presented.

In this work, we propose a third possibility which is alternative to 
unification
but nevertheless maintains the prediction of the weak angle at low energies.
Specifically, we consider  the strong and electroweak interactions to 
arise from two different
collections of coinciding branes, leading to two different gauge couplings,
\cite{twoc}.
Assuming that the low energy spectrum of the (non-supersymmetric) Standard Model
can be derived by a type I/I$^\prime$ string vacuum, the normalization of the
hypercharge is determined in terms of the two gauge couplings and leads
naturally to the right value of $\sin^2\theta_W$ for a string scale of the order
of a few TeV. The electroweak gauge symmetry is broken by the vacuum expectation
values of two Higgs doublets, which are both necessary in the present context to
give masses to all quarks and leptons.

Another issue of this class of models with TeV string scale is to understand
proton stability. In the model presented here, this is achieved by the
conservation of the baryon number which turns out to be a perturbatively
exact global symmetry, remnant of an anomalous $U(1)$ gauge symmetry broken by
the Green-Schwarz mechanism. Specifically, the anomaly is canceled by shifting a
corresponding axion field that gives mass to the $U(1)$ gauge boson.

\section*{The model and the weak angle}

The gauge group closest to the $SU(3)\times SU(2)\times U(1)$ of the Standard
Model one can hope to derive from type I/I$^\prime$ string theory in the above
context is $U(3)\times U(2)\times U(1)$. The first factor arises from three
coincident D-branes (``color" branes). An open string with one end on them is a
triplet under $SU(3)$ and carries the same $U(1)$ charge for all three components.
Thus, the $U(1)$ factor of $U(3)$ has to be identified with {\it gauged} baryon
number. Similarly, $U(2)$ arises from two coincident ``weak" D-branes and the
corresponding abelian factor is identified with {\it gauged} weak-doublet
number. A priori, one might expect that $U(3)\times U(2)$ would be the minimal
choice. However, this is not good enough because the hypercharge cannot be
expressed as a linear combination of baryon and weak-doublet numbers~\footnote{See
nevertheless the comments at the end of this section for a string embedding of the
Standard Model based on $U(3)\times U(2)$, where the two
$U(1)$'s are not the baryon and weak-doublet numbers. The model is though
unsatisfactory for phenomenological reasons.}. Therefore,
at least one additional $U(1)$ factor corresponding to an extra D-brane
(``$U(1)$" brane) is necessary in order to accommodate the Standard 
Model. In principle this $U(1)$ brane can be chosen to be independent of the other
two collections with its own gauge coupling. To improve the predictability of the
model, here we choose to put it on top of either the color or the weak D-branes. In
either case, the model has two independent gauge couplings
$g_3$ and $g_2$ corresponding, respectively, to the gauge groups $U(3)$ and
$U(2)$. The $U(1)$ gauge coupling $g_1$ is equal to either $g_3$ or $g_2$.

Let us denote by $Q_3$, $Q_2$ and $Q_1$ the three $U(1)$ charges of $U(3)\times
U(2)\times U(1)$, in a self explanatory notation. Under $SU(3)\times SU(2)\times
U(1)_3\times U(1)_2\times U(1)_1$, the members of a family of quarks and
leptons have the following quantum numbers:
\bea
&Q &({\bf 3},{\bf 2};1,w,0)_{1/6}\nonumber\\
&u^c &({\bf\bar 3},{\bf 1};-1,0,x)_{-2/3}\nonumber\\
&d^c &({\bf\bar 3},{\bf 1};-1,0,y)_{1/3}\label{charges}\\
&L   &({\bf 1},{\bf 2};0,1,z)_{-1/2}\nonumber\\
&l^c &({\bf 1},{\bf 1};0,0,1)_1\nonumber
\eea
Here, we normalize all $U(N)$ generators according to
${\rm Tr}T^aT^b=\delta^{ab}/2$, and measure the corresponding $U(1)_N$ charges
with respect to the coupling $g_N/\sqrt{2N}$, with $g_N$ the $SU(N)$ coupling
constant. Thus, the fundamental representation of $SU(N)$ has $U(1)_N$ charge
unity. The values of the $U(1)$ charges $x,y,z,w$ will be fixed below so that
they lead to the right hypercharges, shown for completeness as subscripts.

The quark doublet $Q$ corresponds necessarily to a massless excitation of an
open string with its two ends on the two different collections of branes. The
$Q_2$ charge $w$ can be either $+1$ or $-1$ depending on whether $Q$
transforms as a $\bf 2$ or a $\bf\bar 2$ under $U(2)$. The antiquark $u^c$
corresponds to fluctuations of an open string with one end on the color
branes and the other on the $U(1)$ brane for $x=\pm 1$, or on other branes in
the bulk for $x=0$. Ditto for $d^c$. Similarly, the lepton doublet $L$
arises from an open string with one end on the weak branes and the other
on the $U(1)$ brane for $z=\pm 1$, or in the bulk for $z=0$. Finally, $l^c$
corresponds necessarily to an open string with one end on the $U(1)$ brane and
the other in the bulk. We defined its $Q_1=1$.

The weak hypercharge $Y$ is a linear combination of the three
$U(1)$'s:\footnote{A study of hypercharge embeddings in gauge groups obtained
from M-branes was considered in Ref. \cite{west}.
In the context of Type I groundstates such embeddings were considered in 
\cite{ib}.}
\be
Y=c_1 Q_1+c_2 Q_2+c_3 Q_3\, .
\label{Y}
\ee
$c_1=1$ is fixed by the charges of $l^c$ in eq.~(\ref{charges}), while
for the remaining two coefficients and the unknown charges $x,y,z,w$, we obtain
four possibilities:
\bea
c_2 =-{1\over 2}\, ,\, c_3=-{1\over 3}\, ;&
x=-1\, ,\, y=0\, ,\, z=0\, ,\, w=-1\nonumber\\
c_2 ={1\over 2}\, ,\, c_3=-{1\over 3}\, ;&
x=-1\, ,\, y=0\, ,\, z=-1\, ,\, w=1\nonumber\\
c_2 =-{1\over 2}\, ,\, c_3={2\over 3}\,  ;&\!\!\!\!\!\!\!\!\!
x=0\, ,\, y=1\, ,\, z=0\, ,\, w=1\label{solutions}\\
c_2 ={1\over 2}\, ,\, c_3={2\over 3}\, ;&
x=0\, ,\, y=1\, ,\, z=-1\, ,\, w=-1\nonumber
\eea
Orientifold models realizing the $c_3=-1/3$ embedding in the supersymmetric case
with intermediate string scale $M_s\sim 10^{11}$ GeV have been described in 
\cite{ib}.

To compute the weak angle $\sin^2\theta_W$, we use from eq.~(\ref{Y}) that the
hypercharge coupling $g_Y$ is given by~\footnote{The gauge couplings
$g_{2,3}$ are determined at the tree-level by the string coupling and other
moduli, like radii of longitudinal dimensions. In higher orders, they also
receive string threshold corrections.}:
\be
{1\over g_Y^2}={2\over g_1^2}+{4c_2^2\over g_2^2}+
{6c_3^2\over g_3^2}\, ,
\label{gY}
\ee
with $g_1=g_2$ or $g_1=g_3$ at the string scale. On the other hand, with the
generator normalizations employed above, the weak $SU(2)$ gauge coupling is
$g_2$. Thus,
\be
\sin^2\theta_W\equiv{g_Y^2\over g_2^2+g_Y^2}=
{1\over 1+4c_2^2+2g_2^2/g_1^2+6c_3^2g_2^2/g_3^2}\, ,
\label{sintheta}
\ee
which for $g_1=g_2$ reduces to:
\be
\sin^2\theta_W(M_s)=
{1\over 4+6c_3^2g_2^2(M_s)/g_3^2(M_s)}\, ,
\label{sintheta12}
\ee
while for $g_1=g_3$ it becomes:
\be
\sin^2\theta_W(M_s)=
{1\over 2+2(1+3c_3^2)g_2^2(M_s)/g_3^2(M_s)}\, .
\label{sintheta13}
\ee

We now show that the above predictions agree with the experimental value for
$\sin^2\theta_W$ for a string scale in the region of a few TeV. For this
comparison, we use the evolution of gauge couplings from the weak scale $M_Z$ as
determined by the one-loop beta-functions of the Standard Model with three
families of quarks and leptons and one Higgs doublet,
\be
{1\over \alpha_i(M_s)}={1\over \alpha_i(M_Z)}-
{b_i\over 2\pi}\ln{M_s\over M_Z}\ ; \quad i=3,2,Y
\ee
where $\alpha_i=g_i^2/4\pi$ and $b_3=-7$, $b_2=-19/6$, $b_Y=41/6$. We also use
the measured values of the couplings at the $Z$ pole 
$\alpha_3(M_Z)=0.118\pm 0.003$, $\alpha_2(M_Z)=0.0338$, $\alpha_Y(M_Z)=0.01014$
(with the errors in $\alpha_{2,Y}$ less than 1\%) \cite{data}. 

In order to compare the theoretical relations for the two cases
(\ref{sintheta12}) and (\ref{sintheta13}) with the experimental value of
$\sin^2\theta_W=g_Y^2/(g_2^2+g_Y^2)$ at $M_s$, we plot in Fig.~1
the corresponding curves as functions of $M_s$. 
%
\begin{figure}[htb]
\epsfxsize=6.5in
\epsfysize=5in
\hspace{1.5cm}
\epsffile{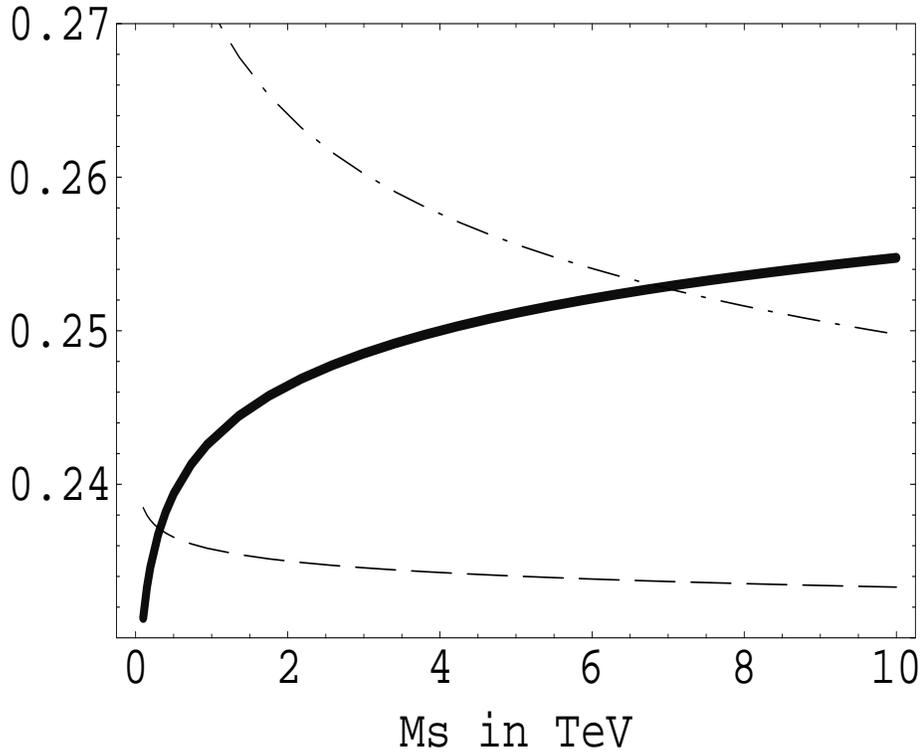}
\vspace{-2.5cm}
\caption{The experimental value of $\sin^2\theta_W$ (thick curve), together with
the theoretical predictions (\ref{sintheta12}) with $c_3=-1/3$ (dashed line) and
(\ref{sintheta13}) with $c_3=2/3$ (dotted-dashed), are plotted as functions of 
the string scale $M_s$.}
\label{sin}
\end{figure}
The solid line is the experimental curve. The dashed line is the plot of the
function (\ref{sintheta12}) for $c_3=-1/3$ while the dotted-dashed line
corresponds to the function (\ref{sintheta13}) for $c_3=2/3$. Thus, the second
case, where the $U(1)$ brane is on top of the color branes, is compatible with
low energy data for $M_s\sim 6-8$ TeV. This selects the last two possibilities of
charge assignments in Eq.~(\ref{solutions}). Indeed, the curve corresponding to
$g_1=g_3$ and $c_3=-1/3$ intersects the experimental curve for $\sin^2\theta_W$
at a scale $M_s$ of the order of a few thousand TeV. Since this value appears to
be too high to protect the hierarchy, it is less interesting and is not shown in
Fig.~1. The other case, where the $U(1)$ brane is on top of the weak branes, is
not interesting either. For $c_3=2/3$, the corresponding curve does not intersect
the experimental one at all and is not shown in the Fig.~1, while the case of
$c_3=-1/3$ leads to $M_s$ of a few hundred GeV and is most likely excluded
experimentally. In the sequel we shall restrict ourselves to the last two
possibilities of Eq.~(\ref{solutions}).

{}From the general solution (\ref{solutions}) and the requirement that the Higgs
doublet has hypercharge $1/2$, one finds the following possible assignments for
it, in the notation of Eq.~(\ref{charges}):
\bea
\label{H}
c_2=-{1\over 2}\ :\qquad &H\ \ ({\bf 1},{\bf 2};0,1,1)_{1/2}\quad
&H'\ \ ({\bf 1},{\bf 2};0,-1,0)_{1/2}\\
c_2={1\over 2}\ :\qquad &{\tilde H}\ \ ({\bf 1},{\bf 2};0,-1,1)_{1/2}\quad
&{\tilde H}'\ \ ({\bf 1},{\bf 2};0,1,0)_{1/2}
\label{Htilde}
\eea
It is straightforward to check that the allowed (trilinear) Yukawa terms are:
\bea
\label{HY}
c_2=-{1\over 2}\ :\qquad H'Qu^c\ ,\quad H^\dagger Ll^c
\ ,\quad H^\dagger Qd^c\\
c_2={1\over 2}\ :\qquad {\tilde H}' Qu^c\ ,\quad {\tilde H}'^\dagger Ll^c
\ ,\quad {\tilde H}^\dagger Qd^c
\label{HtildeY}
\eea
Thus, two Higgs doublets are in each case necessary and sufficient to give masses
to all quarks and leptons. Let us point out that the presence of the second Higgs
doublet changes very little the curves of Fig.~1 and consequently our previous
conclusions about $M_s$ and $\sin^2\theta_W$.

A few important comments are now in order:

\noindent
(i) The spectrum we assumed in Eq.~(\ref{charges}) does not contain right-handed
neutrinos on the branes. They could in principle arise from open strings in the
bulk. Their interactions with the particles on the branes would then be
suppressed by the large volume of the transverse space \cite{Rnus}. 
More specifically, conservation of the three U(1) charges allow for the
following Yukawa couplings involving the right-handed neutrino $\nu_R$:
\be
c_2=-{1\over 2}\;:\;\;\; H'~L~{\nu_L}\;\;\;\;\;\;;\;\;\;\;\;\; c_2={1\over 2}\;:\;\;\;
\tilde H~ L ~\nu_R
\ee
These couplings lead to Dirac type neutrino masses between $\nu_L$ from $L$ and
the zero mode of $\nu_R$, which is naturally supressed by the volume of the
bulk.
 
\noindent
(ii) Implicit in the above was our assumption of three generations (\ref{charges})
of quarks and lepton in the light spectrum. They can arise, for example, from
an orbifold action along the lines of the model described in Ref.~\cite{ib}.

\noindent
(iii) From Eq.~(\ref{sintheta13}) and Fig.~1, we find the ratio of the $SU(2)$ and
$SU(3)$ gauge couplings at the string scale to be $\alpha_2/\alpha_3\sim 0.4$. This
ratio can be arranged by an appropriate choice of the relevant moduli. For
instance, one may choose the weak branes to extend along one extra dimension
transverse to the color branes, with size around twice the string length. Another
possibility would be to move slightly off the orientifold point which may be
necessary also for other reasons (see discussion towards the end of the paper).

\noindent
(iv) Finally, it should be stressed that the charge assignments (\ref{charges})
were based on the assumption that the anti-quarks $u^c$ and $d^c$ arise as
excitations of open strings with only one end on the color D-branes. This is not
however the only possibility. The fact that the ${\bf\bar 3}$ of $SU(3)$ can also
be obtained as the antisymmetric product of two ${\bf 3}$'s implies that $u^c$ and
$d^c$ may also arise as fluctuations of open strings with both ends on the color
branes. Similarly, the anti-lepton $l^c$ which is $SU(2)$ singlet can be obtained
as the antisymmetric product of two doublets and consequently it may arise as a
fluctuation of an open string with both ends on the weak branes. In these cases,
the quantum numbers of the corresponding particles will be:
\be
u'^c\ :\ ({\bf\bar 3},{\bf 1};2,0,0)_{-2/3}\qquad
d'^c\ :\ ({\bf\bar 3},{\bf 1};2,0,0)_{1/3}\qquad
l'^c\ :\ ({\bf 1},{\bf 1};0,\mp 2,0)_1
\ee
One should then repeat the previous analysis from the beginning, with any
combination of the particles $u^c$, $d^c$ and $l^c$ in Eq.~(\ref{charges}) replaced
by the corresponding $u'^c$, $d'^c$ and $l'^c$. However, as we argue next, the only
physically viable alternative scenario to the one discussed above is just to
replace $l^c$ by $l'^c$.

First observe that $d^c$ cannot be replaced by $d'^c$. Indeed, this would fix 
$c_3=1/6$ and the $Q$ hypercharge would set $c_2=0$. It is then easy to see
that one cannot satisfy the hypercharge assignments of leptons for either
choice of $l^c$ or $l'^c$. Next, let us replace $u^c$ by $u'^c$. This fixes
$c_3=-1/3$. If we keep $l^c$, then $c_1=1$ and one is left with the first
two lines of Eq.~(\ref{solutions}) as the two possible solutions for $y,z,w$
($x$ is absent in this case). From our previous analysis of $\sin^2\theta_W$,
these solutions are not very interesting since the string scale comes out to
be either too low or too high. On the other hand, if we substitute also $l^c$
by $l'^c$, the solutions for the remaining parameters are: (a) the second
line of Eq.~(\ref{solutions}) with $c_1=1$ as before, which is uninteresting,
and (b) the first line of Eq.~(\ref{solutions}) with $c_1$ undetermined.
In this case, the $Q_1$ charges of all particles vanish, the corresponding
gauge field decouples, and the gauge group becomes effectively 
$U(3)\times U(2)$ with $Y=-Q_2/2-Q_3/3$. At first sight, this seems
to be a more economical embedding of the Standard Model than the one
based on $U(3)\times U(2)\times U(1)$. In this case, the $g_1$ term
drops from Eq.~(\ref{gY}) and the weak angle is given by
$1/\sin^2\theta_W(M_s)=2+2g_2^2(M_s)/3g_3^2(M_s)$.
Unfortunately, comparison with the experimental value of $\sin^2\theta_W$ 
at $M_Z$ requires a string scale of the order of $10^{14}$ GeV. An additional
unsatisfactory feature of the models obtained by replacing $u^c$ with $u'^c$ 
is the absence of appropriate Yukawa couplings to give masses to the up-quarks.

The last case to be examined is the substitution of $l^c$ alone by $l'^c$.
This leads to the same four solutions (\ref{solutions}) as with $l^c$, and
thus, to the same conclusions for $\sin^2\theta_W$ and $M_s$. However, 
the case with $c_2=1/2$ is problematic because the charge
assignments do not allow tree-level Yukawa interactions to give masses
to the leptons. In the case with $c_2=-1/2$, the Yukawa couplings 
of the leptons (\ref{HtildeY}) are slightly modified to
\be
c_2=-{1\over 2}\ :\qquad H'^\dagger Ll'^c\, ,
\ee
implying that they acquire masses from the Higgs which gives masses also to the
up-quarks.

\section*{Extra $U(1)$'s, anomalies and proton stability}

The model under discussion has three $U(1)$ gauge interactions corresponding to
the generators $Q_1$, $Q_2$, $Q_3$. From the previous analysis, the hypercharge
was shown to be either one of the two linear combinations:
\be
Y=Q_1\mp{1\over 2} Q_2+{2\over 3}Q_3\, .
\ee 
It is easy to see that the remaining two $U(1)$ combinations orthogonal to $Y$ are
anomalous. Indeed, the generic two-parameter generator orthogonal to $Y$ is
\be
{\tilde Q}=(\pm{\beta\over 2}-{2\gamma\over 3})Q_1+\beta Q_2 +\gamma Q_3\, ,
\label{Qtilde}
\ee
which satisfies ${\rm Tr}{\tilde Q}T_{SU(2)}^2=\pm 5\beta/4-\gamma/3$ and 
${\rm Tr}{\tilde Q}T_{SU(3)}^2=2\beta +3\gamma/2$ (for $c_2=-1/2$), or
$-3\beta/4+11\gamma/6$ (for $c_2=1/2$); they can both vanish only for
$\beta=\gamma=0$.

We have assumed throughout that this model can be derived as a consistent type
I/I$^\prime$ string vacuum without additional light states charged under
$U(3)\times U(2)\times U(1)$. In such a vacuum, the anomalies should be canceled by
appropriate shifts of Ramond-Ramond axions in the bulk \cite{gs}. 
Since we have two independent anomalous $U(1)$ currents, we need two axions $a,
a'$ that couple to the non-abelian Pontryagin densities with coefficients fixed
by the anomalies. The relevant part of the
low-energy effective lagrangian can be written as:
\bea
\label{Leff}
{\cal L}_{eff}^{(1)} &=& {1\over 2}(\partial a+\lambda M_s A)^2+
{1\over 32\pi^2}{a\over \lambda M_s}
(k_2{\rm Tr}F_2\wedge F_2+k_3{\rm Tr}F_3\wedge F_3) \\
&+&{1\over 2}(\partial a'+\lambda M_s A')^2+
{1\over 32\pi^2}{a'\over \lambda M_s}
(k'_2{\rm Tr}F_2\wedge F_2+k'_3{\rm Tr}F_3\wedge F_3) \, ,
\nonumber
\eea
where $F_2$ and $F_3$ are the $SU(2)$ and $SU(3)$ gauge field strengths, and
$A$, $A'$ the gauge fields corresponding to two independent and mutually
orthogonal anomalous abelian charges $Q_A$, $Q_{A'}$ of the form
(\ref{Qtilde}). $k_2$, $k'_2$ ($k_3$, $k'_3$) are their respective mixed
anomalies with $SU(2)$ ($SU(3)$) given by
\be
k_2={\rm Tr}Q_AT_{SU(2)}^2\ ,\quad k'_2={\rm Tr}Q_{A'}T_{SU(2)}^2\ ,\quad
k_3={\rm Tr}Q_AT_{SU(3)}^2\ ,\quad k'_3={\rm Tr}Q_{A'}T_{SU(3)}^2\ ,
\ee
while $\lambda$ is a calculable parameter in every particular string vacuum. The
theory is invariant under the gauge transformation $A\to A +\partial\Lambda/g_A$,
$a\to a-\lambda M_s \Lambda$, together with appropriate transformations of the
fermion fields. Indeed, under this transformation the action (\ref{Leff}) changes by
exactly the amount necessary to cancel the phase of the fermionic determinant.
Ditto for $A'$. According to Eq.~(\ref{Leff}), the gauge fields
$A$ and $A'$ become massive with masses  $\lambda g_A M_s$ and $\lambda g_{A'}
M_s$, respectively, with $g_A$ and $g_{A'}$ the corresponding gauge couplings.
The axions $a$ and $a'$ become their longitudinal components. Note that we have
chosen $A$ and $A'$ so that their mass matrix is diagonal. Gravitational anomalies
proportional to the trace of a single charge are also canceled by similar axionic
couplings to $R\wedge R$.

This mechanism can be generalized to show the cancellation of the mixed $U(1)$
anomalies. These are of two types. First, the ones associated to the
non-vanishing traces ${\rm Tr}Q_AY^2\equiv k_Y$ and 
${\rm Tr}Q_{A'}Y^2\equiv k'_Y$ can be canceled by introducing in 
${\cal L}_{eff}^{(1)}$ the additional terms
\be
{\cal L}_{eff}^{(1)}\to {\cal L}_{eff}^{(1)}+{1\over 32\pi^2}{1\over \lambda M_s}
(k_Ya+k'_Ya')F_Y\wedge F_Y\, ,
\ee
which are needed to cancel the $F_Y\wedge F_Y$ contribution to the divergence
of the two anomalous currents. The coefficients $k_Y$ and $k'_Y$ can be deduced
from the anomaly of the generic current (\ref{Qtilde}). In the case of $l^c$ we
obtain ${\rm Tr}{\tilde Q}Y^2=4\beta/3-43\gamma/18$ (for $c_2=-1/2$), or
$-\beta/12-37\gamma/18$ (for $c_2=1/2$), while for $l'^c$ (and $c_2$ necessarily
$-1/2$) ${\rm Tr}{\tilde Q}Y^2=-7\beta/6-31\gamma/18$.

Second, there are mixed anomalies related to the non-vanishing trace
${\rm Tr}Y{\tilde Q}^2=\beta^2/2+20\gamma^2/9-\beta\gamma/3$ (for $c_2=-1/2$),
or $-3\beta^2/4+16\gamma^2/9-5\beta\gamma/3$ (for $c_2=1/2$), or
$17\beta^2/4+16\gamma^2/9+\beta\gamma/3$ (in the case of $l'^c$ for $c_2=-1/2$).
Using this general formula, we can uniquely determine the two orthogonal
combinations $Q_A$ and $Q_{A'}$ in such a way that the triple mixed trace vanishes.
We thus have:
\be
{\rm Tr}YQ_A^2\equiv\xi_A\ ,\quad {\rm Tr}YQ_{A'}^2\equiv\xi_{A'}\ ,\quad
{\rm Tr}YQ_AQ_{A'}=0\, .
\label{YAA}
\ee
These mixed anomalies seem to violate the hypercharge gauge invariance of the
Standard Model. However, in the context of a consistent string theory, they
should also be eliminated. This can be achieved without giving mass to the
hypercharge gauge field $A_Y$ if the low-energy effective lagrangian contains
Chern-Simons terms of the form $A_Y\wedge\omega_A$ and $A_Y\wedge\omega_{A'}$.
Finally, the violation of the $U(1)_A$ and $U(1)_{A'}$ gauge invariances
introduced by these new terms can be remedied by adding non-diagonal axionic
couplings proportional to $aF_Y\wedge F_A$ and $a'F_Y\wedge F_{A'}$. To
summarize, one may cancel all anomalies of our model by modifying the relevant to
the anomaly part of the effective lagrangian ${\cal L}_{eff}^{(1)}$ in
Eq.~(\ref{Leff}) to:
\bea
{\cal L}_{eff}^{\rm anom}={\cal L}_{eff}^{(1)} &+&{1\over 32\pi^2}
{1\over \lambda M_s} (k_Ya+k'_Ya')F_Y\wedge F_Y\\
&-&{1\over 32\pi^2}A_Y\wedge (\xi_A \omega_A+\xi_{A'}\omega_{A'})+
{1\over 32\pi^2}{1\over \lambda M_s}F_Y\wedge (\xi_A aF_A+\xi_{A'}a'F_{A'})
\nonumber\, .
\eea

For completeness, we give the linear combinations $Q_A$ and $Q_{A'}$ that
satisfy Eq.~(\ref{YAA}):
\bea
Q_A &\sim&{3\over 2}Q_1\pm{13\over 3}Q_2+Q_3+t\,(-{2\over 3}Q_1+Q_3)\nonumber\\
Q_{A'}&\sim&-t\,({3\over 2}Q_1\pm{13\over 3}Q_2+Q_3)+
{61\over 4}(-{2\over 3}Q_1+Q_3)
\label{QAs}
\eea
where the $\pm$ signs correspond to $c_2=\mp 1/2$, respectively. For $c_2=1/2$ the
value of $t$ is $t=(1159\pm 13\sqrt{21533})/388$. For $c_2=-1/2$, 
$t=(427\pm 13\sqrt{1342})/54$ for $l^c$, and $t=(-671\pm 91\sqrt{61})/60$ for
$l'^c$.

An important property of the above Green-Schwarz anomaly cancellation mechanism
is that the two $U(1)$ gauge bosons $A$ and $A'$ acquire masses leaving behind
the corresponding global symmetries (\ref{QAs}) \cite{gs}. This is in contrast to
what would had happened in the case of an ordinary Higgs mechanism. These global
symmetries remain exact to all orders in type I string perturbation theory
around the orientifold vacuum. This follows from the topological nature of
Chan-Paton charges in all string amplitudes. On the other hand, one expects
non-perturbative violation of global symmetries and consequently exponentially
small in the string coupling, as long as the vacuum stays at the orientifold point.
Once we move sufficiently far away from it, we expect the violation to become of
order unity. This can be justified in a supersymmetric theory as follows. Every
Ramond-Ramond axion $a$ is part of a chiral superfield $a+im/g_s$ with $g_s$ the
string coupling. Its scalar component $m$ is a NS-NS (Neuveu-Scwharz) closed string
modulus, whose vacuum expectation value (VEV) blows up the orbifold singularities
moving away from the orientifold point. Using the shift of the axion under gauge
transformations, one can form the complex field $e^{(ia-m/g_s)/M_s}$ that
transforms covariantly with charge $-\lambda$. A matter interaction term with
charge $n\lambda$ (with integer $n$), multiplied by the $n$-th power of this field
forms a neutral operator which can appear in the effective action. For
$<m>\ne 0$, one thus obtains charge violating non-perturbative interaction terms
with strength ${\cal O}(e^{-<m>/g_sM_s})$. A small $<m>$ of order $g_sM_s$ leads
therefore to charge violations of order unity.

So, as long as we stay at the orientifold point, all three charges $Q_1$, $Q_2$,
$Q_3$ are conserved and since $Q_3$ is the baryon number, proton stability is
guaranteed.

To break the electroweak symmetry, the Higgs doublets in Eq.~(\ref{H}) or
(\ref{Htilde}) should acquire non-zero VEV's. Since the model is
non-supersymmetric, this may be achieved radiatively \cite{abq}. From
Eqs.~(\ref{HY}) and (\ref{HtildeY}), to generate masses for all quarks and leptons,
it is necessary for both Higgses to get non-zero VEV's. The baryon number
conservation remains intact because both Higgses have vanishing $Q_3$. However, the
linear combination $(t-61/4)Q_A+(t+1)Q_{A'}$ which does not contain $Q_3$, will be
broken spontaneously, as follows from their quantum numbers in Eqs.~(\ref{H}) and
(\ref{Htilde}). This leads to an unwanted massless Goldstone boson of the
Peccei-Quinn type. A possible way out is to break this global symmetry explicitly,
by moving away from the orientifold point along the direction of non-vanishing
$(t-61/4)m+(t+1)m'$, so that baryon number remains conserved.

In conclusion, we presented a particular embedding of the Standard Model in a
non-supersymmetric D-brane configuration of type I/I$^\prime$ string theory. The
strong and electroweak couplings are not unified because strong and weak
interactions live on different branes. Nevertheless, $\sin^2\theta_W$ is naturally
predicted to have the right value for a string scale of the order of a few TeV. The
model contains two Higgs doublets needed to give masses to all quarks and leptons,
and preserves baryon number as a (perturbatively) exact global symmetry. The model
satisfies the main phenomenological requirements for a viable low energy theory
and its explicit derivation from string theory deserves further study.

\section*{Acknowledgments}

This work was partly supported by the EU under TMR contract ERBFMRX-CT96-0090 
and INTAS contract  N 99 0590.
I.A. would like to thank the Department of Physics of the University of Crete,
while T.N.T. the CPHT of the Ecole Polytechnique, as well as the L.P.T. of the
Ecole Normale Sup\'erieure, for their hospitality. E.K. is grateful to C. Bachas,
L. Ib{\'a}{\~n}ez, H.P. Nilles, A. Sagnotti and A. Uranga for discussions.


\end{document}